\documentclass[prx,aps,twocolumn,superscriptaddress]{revtex4}

\usepackage{graphics,amsmath,epsfig}
\usepackage{graphicx,color}
\usepackage{dcolumn}
\usepackage{bm}
\usepackage{amssymb}
\newcommand{\RN}[1]{%
  \textup{\uppercase\expandafter{\romannumeral#1}}%
}

\begin{document}



\title{Frustrated Kondo chains and glassy magnetic phases on the pyrochlore lattice}

\author{Jing Luo}
\affiliation{Department of Physics, University of Virginia, Charlottesville, VA 22904, USA}

\author{Gia-Wei Chern}
\affiliation{Department of Physics, University of Virginia, Charlottesville, VA 22904, USA}

\date{\today}

\begin{abstract}
We present an extensive numerical study of a new type of frustrated itinerant magnetism on the pyrochlore lattice. In this theory, the pyrochlore magnet can be viewed as a cross-linking network of Kondo or double-exchange chains. Contrary to models based on Mott insulators, this itinerant magnetism approach provides a natural explanation for several spin and orbital superstructures observed on the pyrochlore lattice. Through extensive Monte Carlo simulations, we obtain the phase diagrams at two representative electron filling fractions $n = 1/2$ and 2/3. In particular, we show that an intriguing glassy magnetic state characterized by ordering wavevectors $\mathbf q = \left( \frac{1}{3},\frac{1}{3}, 1\right)$ gives a rather satisfactory description of the low temperature phase recently observed in spinel~GeFe$_2$O$_4$.
\end{abstract}

\maketitle

\section{Introduction}

Highly frustrated magnets continue to fascinate physicists with intriguing and sometimes unexpected magnetic phases. This is particularly true for spin systems exhibiting strong geometrical frustration such as pyrochlore antiferromagnets~\cite{moessner06}. Conventionally, frustrated magnets are modeled by the Heisenberg Hamiltonian $\mathcal{H} = \sum_{ij} J_{ij} \mathbf S_i \cdot \mathbf S_j $ within the framework of Mott insulators. For pyrochlore and kagome lattices, the frustrated nearest-neighbor antiferromagnetic spin interactions give rise to a macroscopic ground-state degeneracy~\cite{chalker92,moessner98}. This in turn makes the magnets highly susceptible to small perturbations. Removal of the extensive degeneracy by perturbations beyond $J_1$ leads to unusual spin ordering and even unconventional magnetic phases~\cite{lacroix11,balents10}. For systems with degenerate orbitals, a good starting point is the Kugel-Khomskii Hamiltonian~\cite{kugel73}, which has been successfully employed to understand spin-orbital ordering in frustrated magnets~\cite{tsunetsugu03,matteo04,chern10}. 

Recently, complex spin and/or orbital superstructures observed in spinels such as CuIr$_2$S$_4$~\cite{radaelli02,takubo05}, MgTi$_2$O$_4$~\cite{schmidt04}, and ZnV$_2$O$_4$~\cite{lee04} have posed an intriguing theoretical challenge. Several models have been proposed to explain the experimental results. However, understanding these unusual orderings within the framework of Mott insulators often requires fine tuning or sometimes {\em ad hoc} perturbations. On the other hand, it has been demonstrated in many cases that approaches based on itinerant magnetism provide a very natural explanation for the observed superstructures~\cite{khomskii05,chern11,kato12}. For example, the octamer-order in CuIr$_2$S$_4$ and chiral distortion in MgTi$_2$O$_4$ can be explained as resulting from an orbital driven Peierls instability~\cite{khomskii05,radaelli05}.
Moreover, several of these compounds have been shown to be a bad insulator, indicating that these magnets are in the vicinity of metal-insulator transition~\cite{pardo08,kuntscher12,croft03,wang04}. Recent experiments further support the picture of orbital-Peierls state~\cite{zhou06,yang08}.

The itinerant approach also naturally explains the $\mathbf q = (0, 0, 1)$ magnetic structure of ZnV$_2$O$_4$, which consists of $\uparrow\uparrow\downarrow\downarrow \cdots$ spin chains along $[110]$ directions of the pyrochlore lattice~\cite{chern11,kato12}. Essentially, taking into account the reduced dimensionality of electron hopping in such systems, this interesting commensurate one-dimensional (1D) order can be understood as resulting from the spin-induced nesting instability of 1D Kondo chains. 
Another interesting example is the multiple-$\mathbf q$ magnetic ordering recently observed in spinel GeFe$_2$O$_4$~\cite{zhu16}. At low temperatures, neutron-scattering experiments found diffusive peaks centered at $\mathbf q = (\frac{1}{3}, \frac{1}{3}, 1)$ and other symmetry-related wavevectors, implying a quasi-1D ordering with a tripled unit cell.  Stabilization of this unusual commensurate magnetic order seems rather difficult using the localized spin models. 

In this paper, we present a detailed numerical study of a novel frustrated itinerant spin model for spinel compounds $A B_2 X_4$. In these materials, the octahedral crystal field splits the $3d$ orbitals of the $B$-site magnetic ion into a $t_{2g}$ triplet and a higher energy $e_g$ doublet. Keeping only the dominant $dd\sigma$ transfer integral between the low-energy $t_{2g}$ orbitals, electron hoppings on the pyrochlore lattice can be modeled by a set of one-dimensional (1D) tight-binding chains in this leading order approximation~\cite{chern11}. Inclusion of the on-site Hubbard and Hund's interactions within the mean-field approximation then leads to Kondo or double-exchange type electron-spin couplings. A minimum model is given by a collection of cross-linking Kondo chains running along the $\langle 110 \rangle$ directions of the pyrochlore lattice. Importantly, commensurate 1D spin order can arise naturally as a result of Fermi point nesting instability in Kondo chains with a rational electron filling fraction. A new type of geometrical frustration then results from the fact that the favored 1D spin order cannot be realized on all chains simultaneously, leading to novel 3D magnetic order and to glassy behavior in some cases.


\section{Model and Method}

\begin{figure}
\includegraphics[width = 0.9\columnwidth]{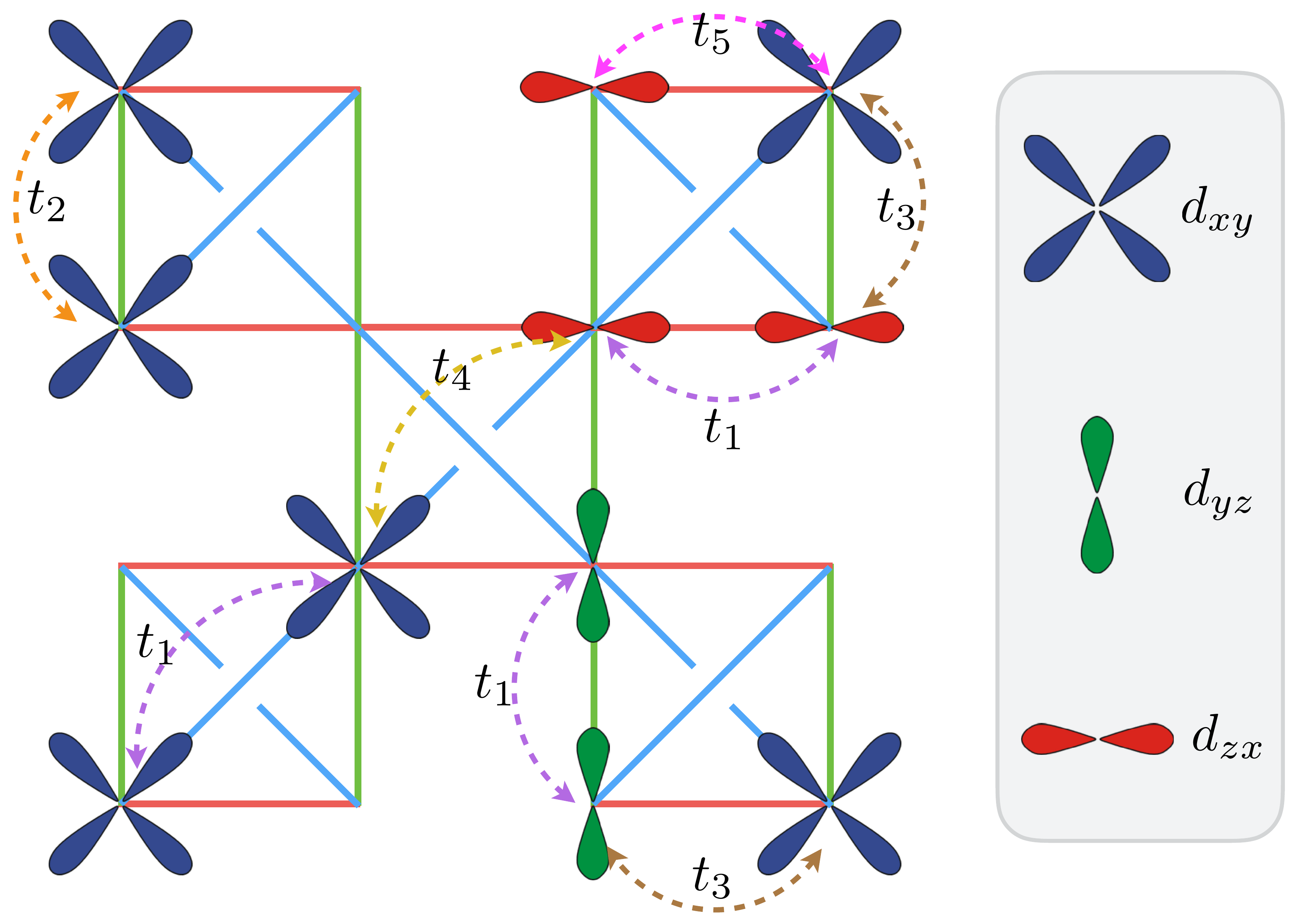}
\caption{\label{fig:hopping} (Color online) The inequivalent transfer integrals between the three $t_{2g}$ orbitals on the pyrochlore lattice: $t_1 = \frac{3}{4} V_{dd\sigma} + \frac{1}{4} V_{dd\delta}$, $t_2 = \frac{1}{2} V_{dd\pi} + \frac{1}{2} V_{dd\delta}$, $t_3 =\frac{1}{2} V_{dd\pi} - \frac{1}{2} V_{dd\delta}$, $t_4 = t_5 = 0$.}
\end{figure}

Our itinerant electron approach to magnetic orders in spinels is based on a mean-field treatment of Hubbard-type Hamiltonian. First, we consider the tight-binding model of $t_{2g}$ orbitals in spinels. As discussed above, the magnetic ions in spinels form a pyrochlore lattice. Fig.~\ref{fig:hopping} shows some representative hopping processes of $t_{2g}$ electrons on the pyrochlore lattice. Here the various hopping integrals are computed using the Slater-Koster formula; the results can be expressed in terms of fundamental bond integrals $V_{dd\sigma}$, $V_{dd\pi}$, and $V_{dd\delta}$~\cite{slater54}. In general, the $\sigma$ bond-integral is much stronger than the $\pi$, and $\delta$ bonds. To the leading-order approximation, we thus neglect contributions from $V_{dd\pi}$ and $V_{dd\delta}$ to the various bond integrals. As a result, only the $t_1$ hopping remains in this approximation, which means only those nearest-neighbor hoppings between the same type of orbitals among appropriate chains dominate, namely,  $d_{xy}$ along $\langle110\rangle, \langle1\bar10\rangle$, $d_{yz}$ along $\langle011\rangle, \langle01\bar1\rangle$ and $d_{zx}$ along $\langle101\rangle, \langle\bar101\rangle$, see Fig.~\ref{fig:hopping}. 

Next we consider the on-site interactions which is dictated by the multi-orbital Hubbard-Kanamori interaction $\mathcal{H}_U$~\cite{kanamori63}. Since we are interested in solutions with non-zero local moment, we use the Hartree-Fock mean-field method to decouple the interaction terms. The mean-field decoupling gives rise to a Kondo-like electron-spin coupling $\mathcal{H}_U = U_{\rm eff} \langle \hat{\mathbf s}_i \rangle \cdot \hat{\mathbf s}_i$, where $\mathbf s_i$ is the electron spin operator, and $U_{\rm eff}$ is an effective Hubbard parameter. For example, for $t_{2g}$ orbitals, $U_{\rm eff} = 4 (U/9 + 4 J_H/9)$, where $U$ and $J_H$ are the on-site Hubbard repulsion and Hund's coupling, respectively. In the case of GeFe$_2$O$_4$, the magnetic Fe$^{2+}$ ions have a $t_{2g}^4\,e_g^2$ electron configuration. Due to strong intra-orbital Hubbard interaction and Hund's coupling, the two $e_g$ electrons remain in the correlated $S = 1$ state. The remaining $t_{2g}^4$ electrons thus form conduction band with a filling fraction $n = 2/3$.

\begin{figure}[t]
\includegraphics[width = 0.9\columnwidth]{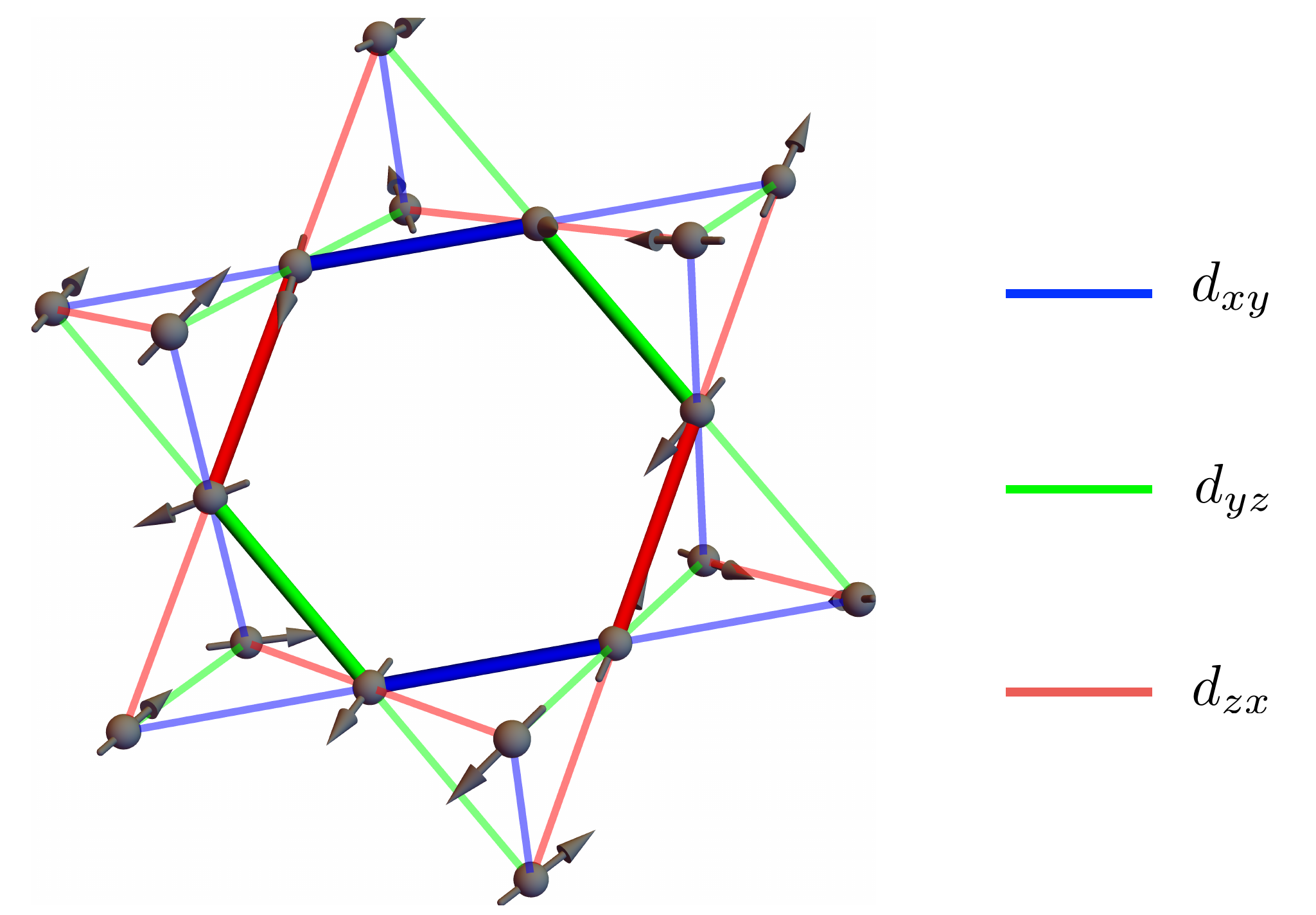}
\caption{\label{fig:t2g} (Color online) Schematic diagram showing the shortest hexagonal loops in the pyrochlore lattice. The three different colors indicate distinct Kondo chains occupied by the three $t_{2g}$ orbitals.}
\end{figure}

We thus arrive at the following Hamiltonian describing cross-linking Kondo chains on the pyrochlore lattice in Fig.~\ref{fig:t2g}:
\begin{eqnarray}
	\label{eq:H_DE}
	\mathcal{H} = -t \sum_{\mu,\sigma} \sum_{\langle ij \rangle \parallel \mu} \left( \hat{c}^{\dagger}_{i\, \mu\sigma} \,\hat{c}^{\;}_{j \mu \sigma} + {\rm h.c.} \right)
	- J \sum_{i, \mu} \mathbf S_i \cdot \hat{\mathbf s}_{i, \mu}
\end{eqnarray}
where $\hat{c}^\dagger_{i, \mu\sigma}$ is the creation operator for electron with spin $\sigma = \uparrow, \downarrow$ and orbital flavor $\mu = xy$, $yz$, $zx$ at site-$i$, $ \langle ij \rangle \parallel \mu $ indicates the nearest-neighbor (NN) pair along the $\langle 110 \rangle$ direction that corresponds to the active $t_{2g}$ orbital $\mu$, the hopping constant $t$ is set to be 1 in all the simulations below, $J \approx U_{\rm eff} \langle \hat{\mathbf s} \rangle$ is the effective Hund's coupling, $\mathbf S_i$ is the $O(3)$ local magnetic moment, and $\hat{\mathbf s}_{i,\mu} = \sum_{\alpha,\beta} c^{\dagger}_{i\mu\alpha}  {\bm{\sigma}_{\alpha\beta}} c_{i\mu\beta}$ is the electron spin operator.

The 1D ferromagnetic Kondo chain, which is the backbone of Hamiltonian~(\ref{eq:H_DE}), have been extensively studied over the years~\cite{tsunetsugu97,garcia04,minami15}. However, the fact that every local spin $\mathbf S_i$ is shared by three Kondo chains introduces competition between different chains. In particular, the cross-linking Kondo chains exhibit a new type of geometrical frustration since the electronic energy of neighboring chains cannot be simultaneously minimized. For example, the shortest hexagonal loops (Fig.~\ref{fig:t2g}) of spins on the pyrochlore lattice contain sites which belong to six different Kondo chains. Consequently, the nearest-neighbor spin-spin correlation favored by individual chains might not be able to extend over the hexagonal loop consistently, leading to frustrated interactions.

Since our main interest is in the potential magnetic orderings of this model, we will assume classical local spins here.  However, even with classical local spins, Monte Carlo simulations of Kondo-lattice models are a challenging task mainly due to the non-local electron-mediated effective interactions between the local moments. Indeed, in the weak-coupling limit $J \ll t$, integrating out the electrons gives rise to a long-range RKKY type spin interactions. For large $J$, one needs to diagonalize the electron tight-binding Hamiltonian that depends on the spin configuration for each Monte Carlo update. For a pyrochlore lattice of linear size $L$, there are $N = 16 L^3$ spins and the dimension of a generic spinful and orbitally degenerate TB Hamiltonian is $D = 2 \times 3 \times N = 96 L^3$. This severely limits the largest accessible lattice sizes, as exact diagonalization scales as $\mathcal{O}(D^3)$ and is computationally very costly.  However, thanks to the 1D nature of the TB model in Eq.~(\ref{eq:H_DE}), each local spin update only requires diagonalizing three chains whose dimension is $D_{\rm 1D} = 4 L$. Specifically, we adopt the standard local Metropolis Monte Carlo method. For a randomly chosen spin, say at site-$i$, we consider rotating the spin from $\mathbf S_i$ to $\mathbf S_i'$.  The energy cost associated with this update comes from the electron energy of the three Kondo chains intersecting at this site, i.e.
\begin{eqnarray}
	\label{eq:dE}
	\Delta E = \sum_{\mu = xy, yz, zx} \left[ \sum_{m=1}^{N_f} \left( \varepsilon_m^{(\mu)}(\mathbf S_i') - \varepsilon_m^{(\mu)}(\mathbf S_i) \right) \right].
\end{eqnarray}
Here $\varepsilon_m^{(\mu)}$ are the eigen-energies of the $\mu$-orbital Kondo chain and $N_f$ is the number of occupied electrons determined by the filling fraction. Once $\Delta E$ is obtained by exactly diagonalizing the three chains intersecting at $\mathbf S_i$, the spin-update is accepted according to the standard Metropolis algorithm with a probability $p_{\rm acc} = {\rm min}[1, \exp(-\Delta E/ k_B T)]$. 
The computational cost of each update thus scales as $\mathcal{O}(D_{\rm 1D}^3) \sim \mathcal{O}(N)$. Each sweep is completed by updating local spins sequentially. The Monte Carlo simulation for the coupled chains is still costly with an overall scaling $\mathcal{O}(N\times D_{\rm 1D}^3) \sim \mathcal{O}(N^2)$, but the efficiency is much improved compared with the full 3D tight-binding model.

\section{Phase diagram}
\begin{figure}[h]
\begin{center}
\includegraphics[width = 0.99\columnwidth]{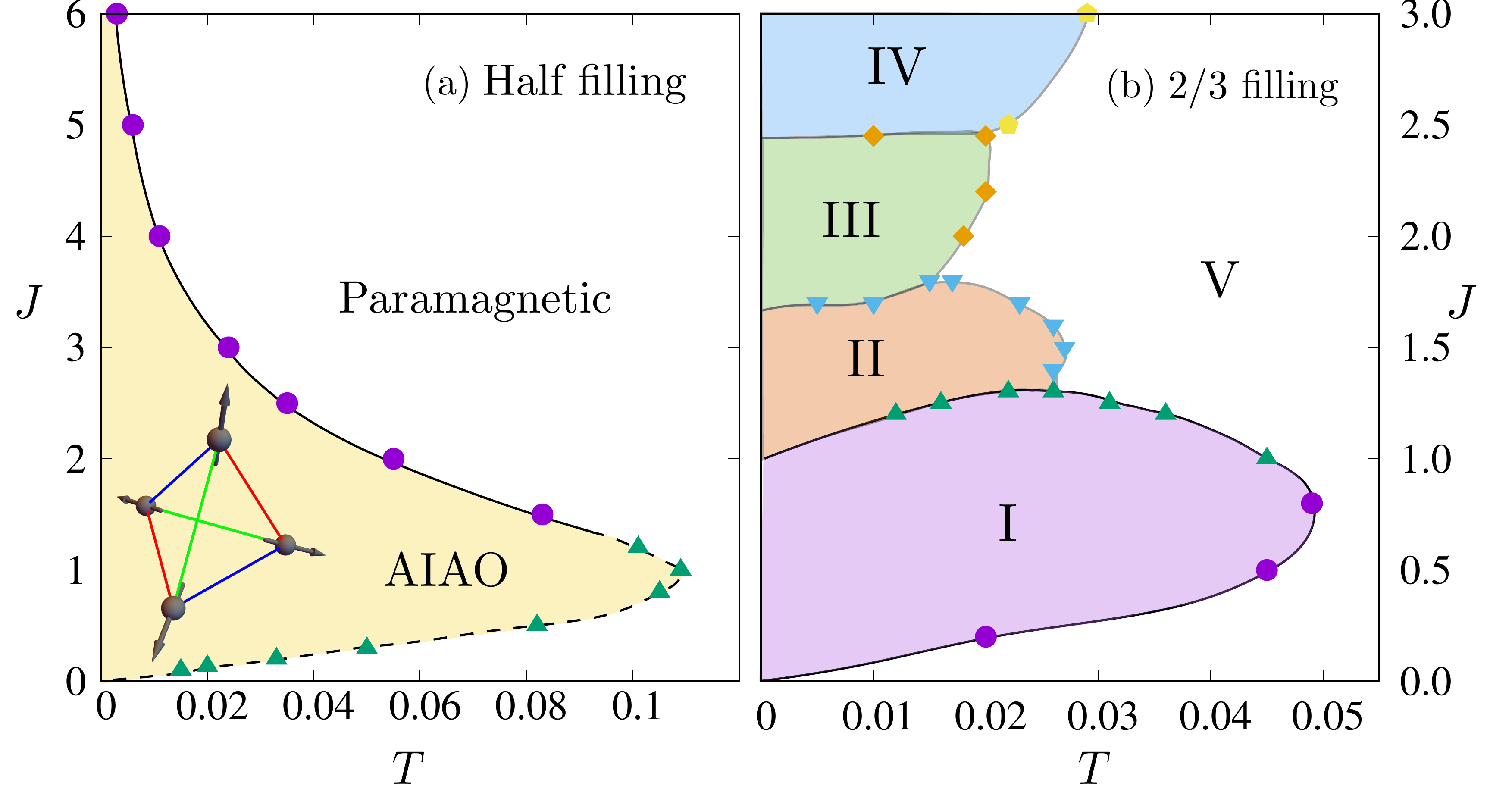}
\caption{
\label{fig:phases}
(Color online) The phase diagrams for (a) half filling and (b) 2/3 filling. Solid and dashed lines represents 1st and 2nd  order phase transitions respectively in both (a) and (b). For half filling, two phases are all-in-all-out phase (AIAO), paramagnetic phase. For 2/3 filling, phases are  ($\RN{1}$) $\mathbf q = \left(\frac{1}{3},\frac{1}{3},1\right)$ order, ($\RN{2}$) $\left(\frac{1}{2}, \frac{1}{2}, \frac{1}{2}\right)$ order, ($\RN{3}$) a unknown magnetic phase characterized by a large spin nematic order parameter, ($\RN{4}$) ferromagnetic phase and ($\RN{5}$) paramagnetic phase.}
\end{center}
\end{figure}

In this part we obtain the phase diagram of Hamiltonian Eq.~(\ref{eq:H_DE}) for two representative filling fractions $n = 1/2$ and $2/3$ based on extensive Monte Carlo simulations; the results are summarized in Fig.~\ref{fig:phases}.

We first discuss the simpler case of half-filling. There is only one ordered phase characterized by the non-coplanar all-in-all-out (AIAO) spin order at low temperatures; see Fig.~\ref{fig:phases}(a). For a half-filled Kondo chain, the nesting of the Fermi points favors a collinear N\'eel order with doubled unit cell, i.e $\uparrow\downarrow\uparrow\downarrow\cdots$. However, it is easy to convince oneself that such collinear ordering cannot be simultaneously realized in the three different chains on the pyrochlore lattice; a manifestation of the geometrical frustration is discussed above. The solution to this conflicted situation is the AIAO order in which an 1D spin-order with a doubled unit cell, albeit with non-collinear spins, still gaps out the Fermi points and lowers the overall energy. 
The AIAO order is characterized by three non-zero staggered magnetization: $\mathbf L_1 = \mathbf S_0 + \mathbf S_1 - \mathbf S_2 - \mathbf S_3$, and the symmetry-related $\mathbf L_2$ and $\mathbf L_3$. Here $\mathbf S_m$ denotes the spin of the $m$-th sublattice (there are four sublattices) of the pyrochlore lattice. A perfect AIAO has $|\mathbf L_1| = |\mathbf L_2| = |\mathbf L_3|$ while their orientations satisfy $\mathbf L_1 \perp \mathbf L_2 \perp \mathbf L_3$. Due to the non-coplanar nature of this magnetic order, the AIAO phase further breaks a $Z_2$ chiral symmetry which is measured by the discrete scalar spin chirality $\chi = \mathbf L_1 \cdot \left( \mathbf L_2 \times \mathbf L_3 \right)$.
The phase boundary of the AIAO order, shown in Fig.~\ref{fig:phases}(a), is determined from the Binder crossing of corresponding staggered order parameters for continuous phase transition at small~$J$.

Interestingly, the transition becomes first-order at large $J$. As in general Kondo-lattice or double-exchange models, the effective Hamiltonian in the large-$J$ limit is given by a Heisenberg model with a dominant NN exchange $J_{\rm AF} \sim t^2/J$. This can be understood as follows. In the $J \to \infty$ limit at half-filling, electrons are localized in individual orbitals of each site with their spins aligned with the local moments. This gives rise to a huge degeneracy which is lifted by the electron hopping. Due to Pauli exclusion principle, electrons can hop to neighboring sites only when their spins are not aligned, thus favoring an antiferromagnetic interaction. Specifically, the effective Hamiltonian corresponds to the energy gain through the second-order process, which is $ E^{(2)}_{ij} \approx - \bigl[t \, \langle \chi_i| \chi_j \rangle \bigr]^2/ J$, where $|\chi_i \rangle$ is the local electron spinor wavefunction. Since Pauli exclusion requires that the spins at $i$ and $j$ must be anti-aligned in order to allow the electrons hop to the NN sites, the inner product of the spinor eigenstates $\langle \chi_i | \chi_j \rangle = \sin(\theta_{ij}/2)$, where $\theta_{ij}$ is the angle between the two local spins. Consequently, we obtain an effective spin interaction: $  E^{(2)}_{ij} = J_{\rm AF} \,\mathbf S_i \cdot \mathbf S_j$ up to a constant, with $J_{\rm AF} \sim t^2/J$.

It is interesting to note that the frustrated nature of the coupled Kondo chains in the large-$J$ limit corresponds to the well known geometrical frustration of AF Heisenberg model on the pyrochlore lattice. The huge ground-state degeneracy of this model leads to a low temperature spin liquid phase. Contrary to the high-temperature paramagnetic phase, disordered spins in this classical spin liquid exhibit strong short-range correlation~\cite{moessner98}. A possible scenario is that the system first enters a correlated classical spin liquid regime at $T \sim J_{\rm AF}$, then undergoes a phase transition at a lower $T_c$ into the AIAO phase. Our detailed analysis shows that the classical spin liquid phase is preempted by the first-order transition, and the system immediately goes to the AIAO phase at a critical~$T_c \sim J_{\rm AF}$.

\begin{figure}[t]
\includegraphics[width = 0.8\columnwidth]{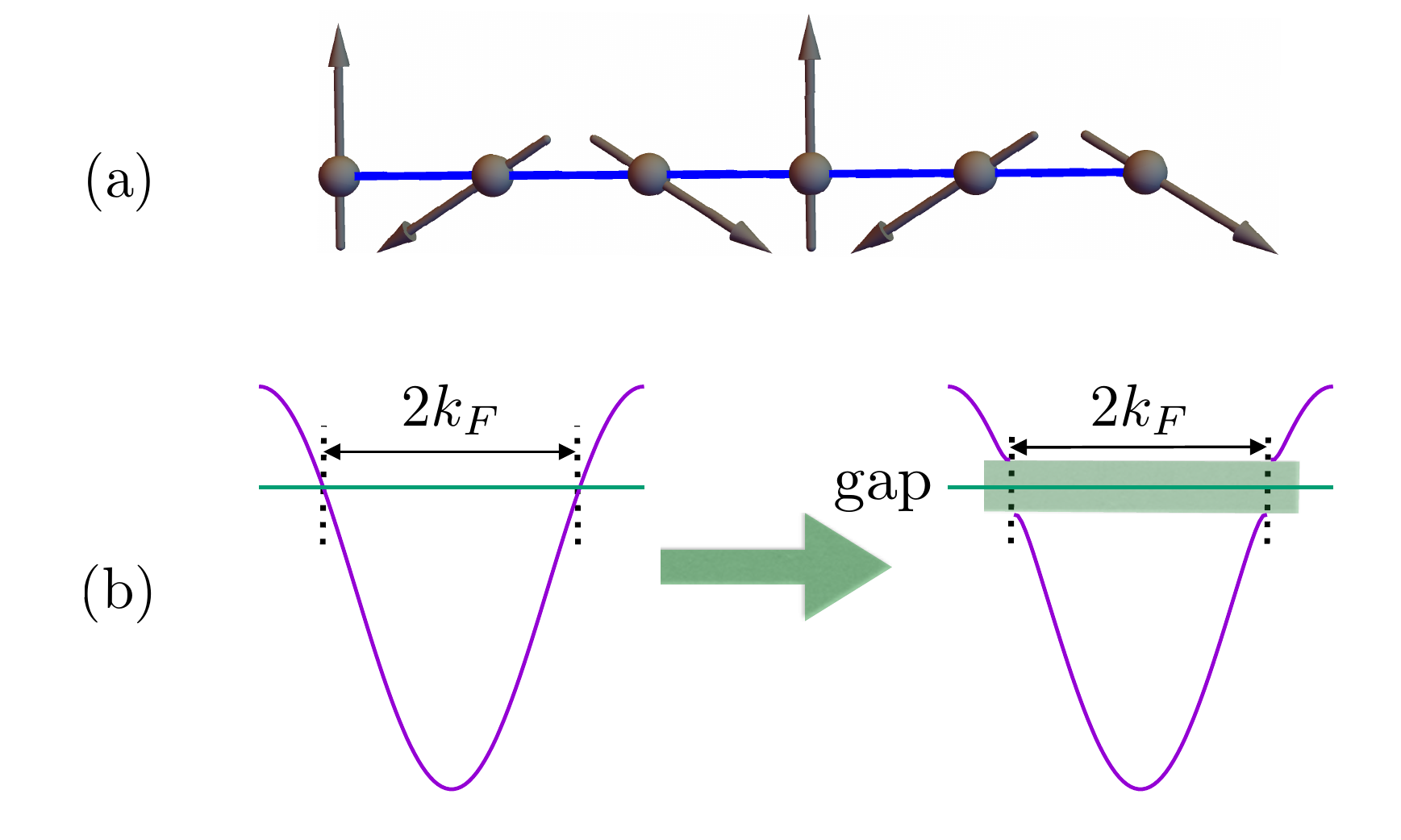}
\caption{\label{fig:nesting} (Color online) (a) The $T \to 0$ ground state of a single Kondo chain. The long-range spin order is characterized by a tripled unit cell with a coplanar almost 120$^\circ$ structure within a unit cell. (b) shows the gap-opening of a $n=2/3$-filled Kondo chain due to Fermi point nesting.}
\end{figure}

We now turn to the case of 2/3-filling. Before discussing the phase diagram of coupled Kondo chains on the pyrochlore lattice, we first consider the ground states of a single Kondo chain. The Fermi wavevector of a 2/3-filled 1D band is $k_F = 2\pi / 3\ell$, where $\ell = \sqrt{2} a/4$ is the 1D lattice constant and $a$ is the size of the cubic unit cell. The system is thus susceptible to perturbations with a wavevector $q = 2 k_F = 4\pi/3\ell$ that gaps out the two Fermi points; see Fig.~\ref{fig:nesting}. Indeed, our Monte Carlo simulations on a single Kondo chain find a magnetic order with a tripled unit cell at $T \to 0$ and small $J$. The three spins within the extended unit cell are coplanar, with a relative angle very close to 120$^\circ$; more details can be found in Appendix~\ref{sec:1D}. 

\begin{figure}[t]
\begin{center}
\includegraphics[width = 0.99\columnwidth]{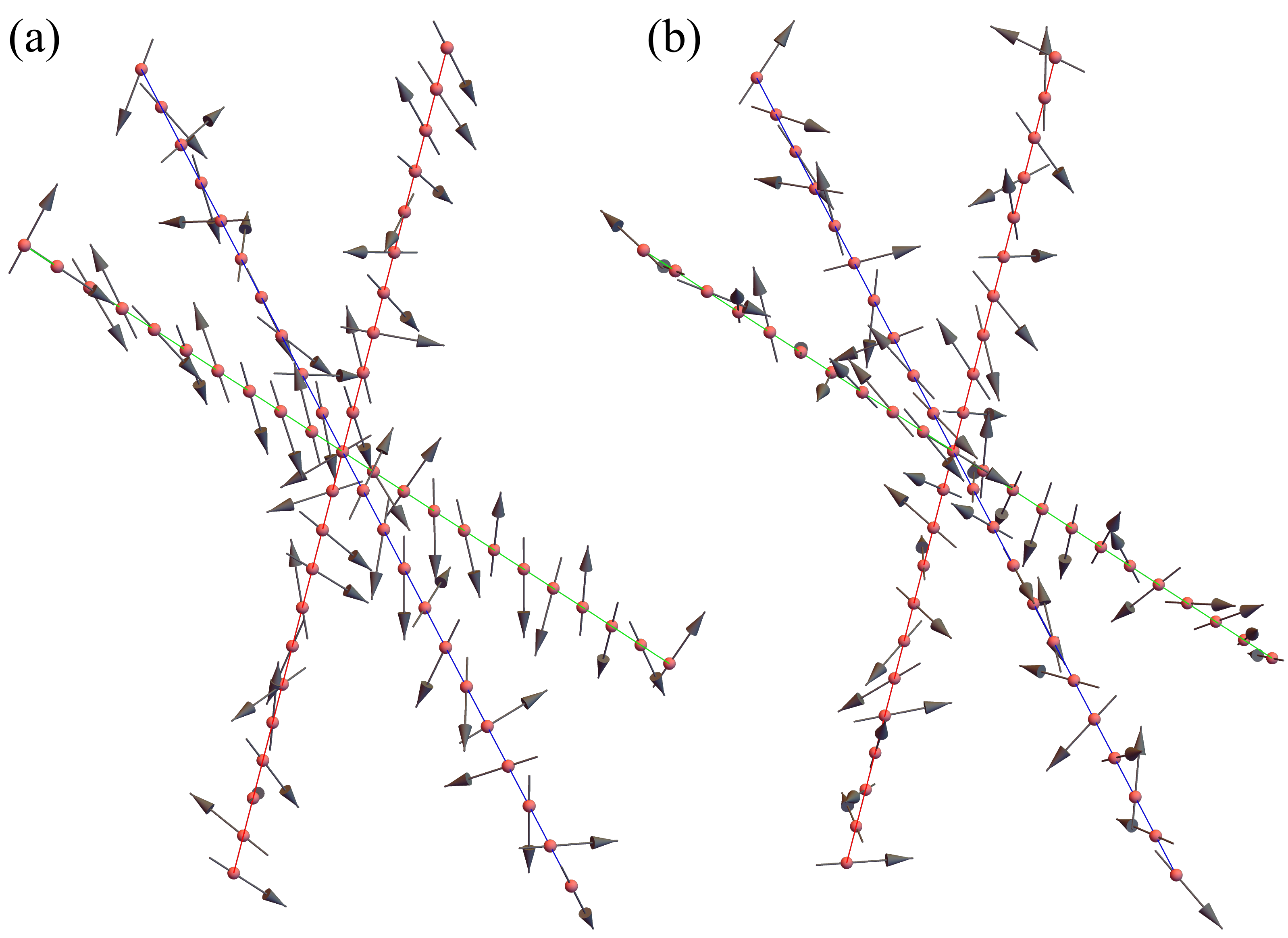}
\caption{
\label{fig:snapshots}
(Color online) Snapshots of the local spin configurations for (a) the $\mathbf q =(\frac{1}{3},\frac{1}{3}, 1)$ order and (b) $\mathbf q =(\frac{1}{2},\frac{1}{2}, \frac{1}{2})$ order.}
\end{center}
\end{figure}

\begin{figure}[b]
\begin{center}
\includegraphics[width = 0.99\columnwidth]{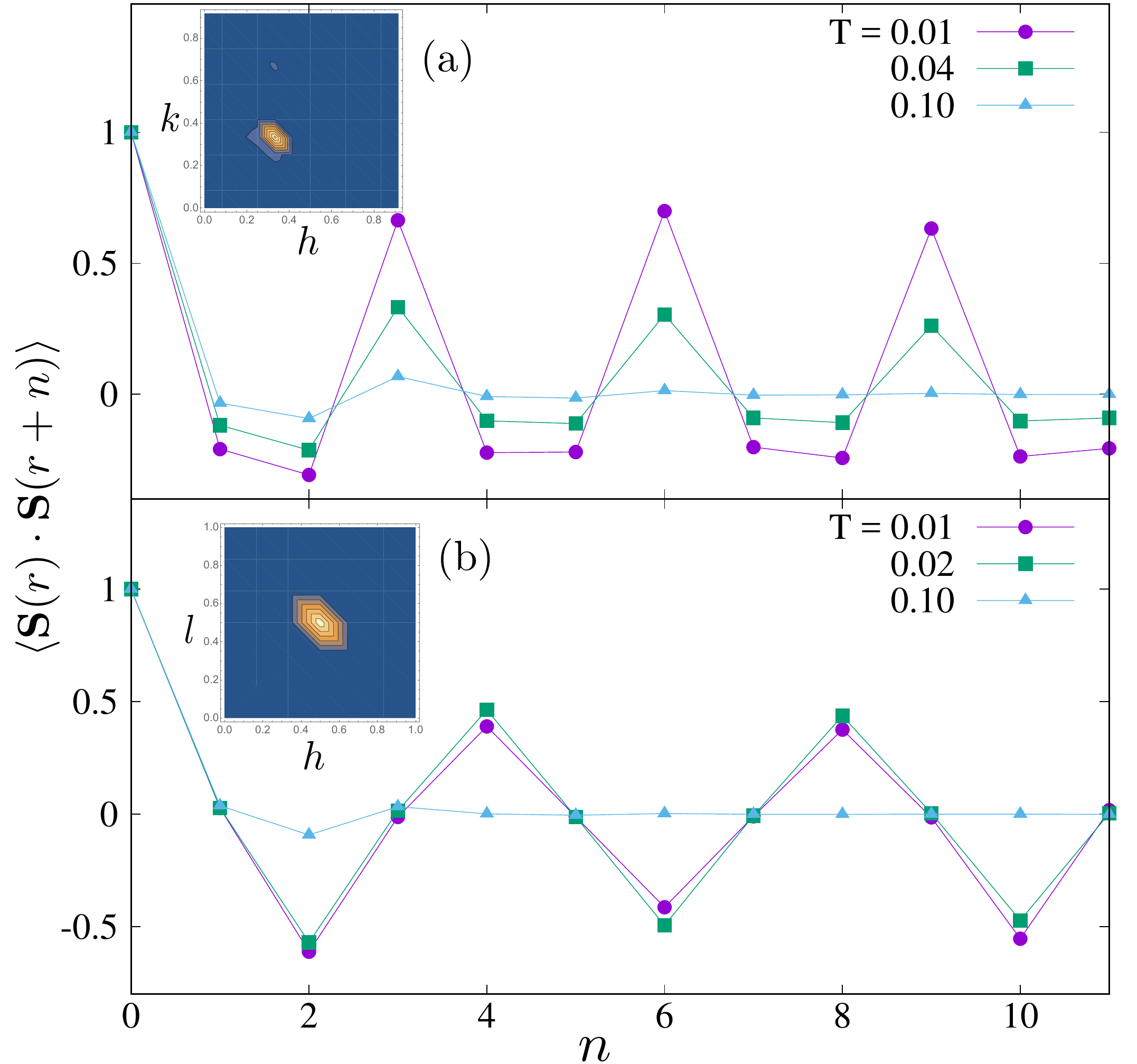}
\caption{
\label{fig:corr}
(Color online) The spin-spin correlation function  $\left< \mathbf S(r)\cdot \mathbf S(r+n) \right>$ averaged over all Kondo chains of the pyrochlore lattice for (a) the $\mathbf q = (\frac{1}{3},\frac{1}{3},1)$ and (b) the $\mathbf q = (\frac{1}{2},\frac{1}{2} ,\frac{1}{2})$ order at $n = 2/3$ filling with (a) $J=1$ for and (b) $J=1.5$. The insets show the corresponding structure factor on the (a) $\mathbf q = (h, k, 1)$ and (b) $\mathbf q = (h, h, l)$ plane.}
\end{center}
\end{figure}

Next we apply the above 1D results to understand the ground-states of coupled Kondo chains in 3D, which is particularly important in explaining the magnetic order of spinel GeFe$_2$O$_4$ where the $t_{2g}$ orbitals are 2/3-filled. From direct inspection of the geometry, one immediately realizes that the above coplanar 1D ground-state cannot be consistently combined in the 3D pyrochlore lattice. This is another manifestation of the geometrical frustration discussed in Fig.~\ref{fig:t2g}. Contrary to the half-filling case, where the frustrated coupling leads to the AIAO long-range order, there is no simple magnetic structure selected in the 2/3-filling case. A snapshot of spin configuration from our Monte Carlo simulations is shown in Fig.~\ref{fig:snapshots}(a). Individual Kondo chains are clearly not in their 1D ground state discussed above. In fact, spins on a given chain are not even coplanar. Although no clear pattern can be seen from this snapshot, detailed characterization shows that a long-range spin-spin correlation with a tripled unit cell nonetheless is developed along each individual chain of the 3D lattice; see Fig.~\ref{fig:corr}(a).  Moreover, the 3D non-coplanar spin order is characterized by multiple wavevectors that are related to $\mathbf q = (\frac{1}{3}, \frac{1}{3}, 1)$ by symmetry, as shown in the inset of Fig.~\ref{fig:corr}(a).

From the phase diagram of single Kondo chain discussed in Appendix~\ref{sec:1D}, the magnetic order at large $J$ cannot be understood from the Fermi-point nesting picture. Here we performed extensive Monte Carlo simulations to obtain the $n=2/3$-filling phase diagram, shown in Fig.~\ref{fig:phases}(b). At small Hund's coupling, the low-$T$ phase is a magnetic order characterized by multiple ordering wavevectors that are related to $\mathbf q = (\frac{1}{3}, \frac{1}{3}, 1)$, as discussed above. Several unusual magnetic structures are obtained at larger $J$. The phase boundaries are mostly first order,  except for the small $J$ regime (purple dots) where the phase transition between paramagnetic and $(\frac{1}{3}, \frac{1}{3}, 1)$-ordered phases might be continuous.

The various 3D phases are loosely related to their 1D counterpart. Upon increasing $J$, the ordering wavevectors first change from $\mathbf q =(\frac{1}{3},\frac{1}{3},1)$ to $(\frac{1}{2},\frac{1}{2},\frac{1}{2})$ at $J \approx t$. The system undergoes another 1st-order transition at $J \approx 1.6 t$ into an unknown magnetic order (phase III) that is characterized by rather large nematic order parameter. We have checked that spins are pretty much frozen in this phase, yet no clear long-range order can be seen from the static structure factor. And finally, the ferromagnetic order takes over as the ground state when $J \gtrsim 2.5 t$.  An interesting case is the $\mathbf q = (\frac{1}{2}, \frac{1}{2}, \frac{1}{2})$ phase at intermediate Hund's coupling $1 \lesssim J \lesssim 1.6$ (phase-II in the phase diagram). A snapshot of local spin configurations on three different chains intersecting at one spin is shown in Fig.~\ref{fig:snapshots}(b). Again, although no clear ordering pattern can be found in the snapshot, detailed analysis showed that individual Kondo chains exhibit a clear 1D spin correlation with a quadrupled unit cell, as shown in Fig.~\ref{fig:corr}(b).  This is in stark contrast to the ground state of a single Kondo chain in the same $J$ regime, where the $T \to 0$ ground state is a multiple-$q$ non-coplanar order. In this case, the ``frustrated" inter-chain coupling actually stabilizes the quadrupled chains and the $\mathbf q = (\frac{1}{2},\frac{1}{2},\frac{1}{2})$ order on the pyrochlore lattice.

\section{Quasi-degeneracy and glassy behaviors of the $\mathbf q = (\frac{1}{3}, \frac{1}{3}, 1)$ phase}

\begin{figure}[t]
\begin{center}
\includegraphics[width = 1.0\columnwidth]{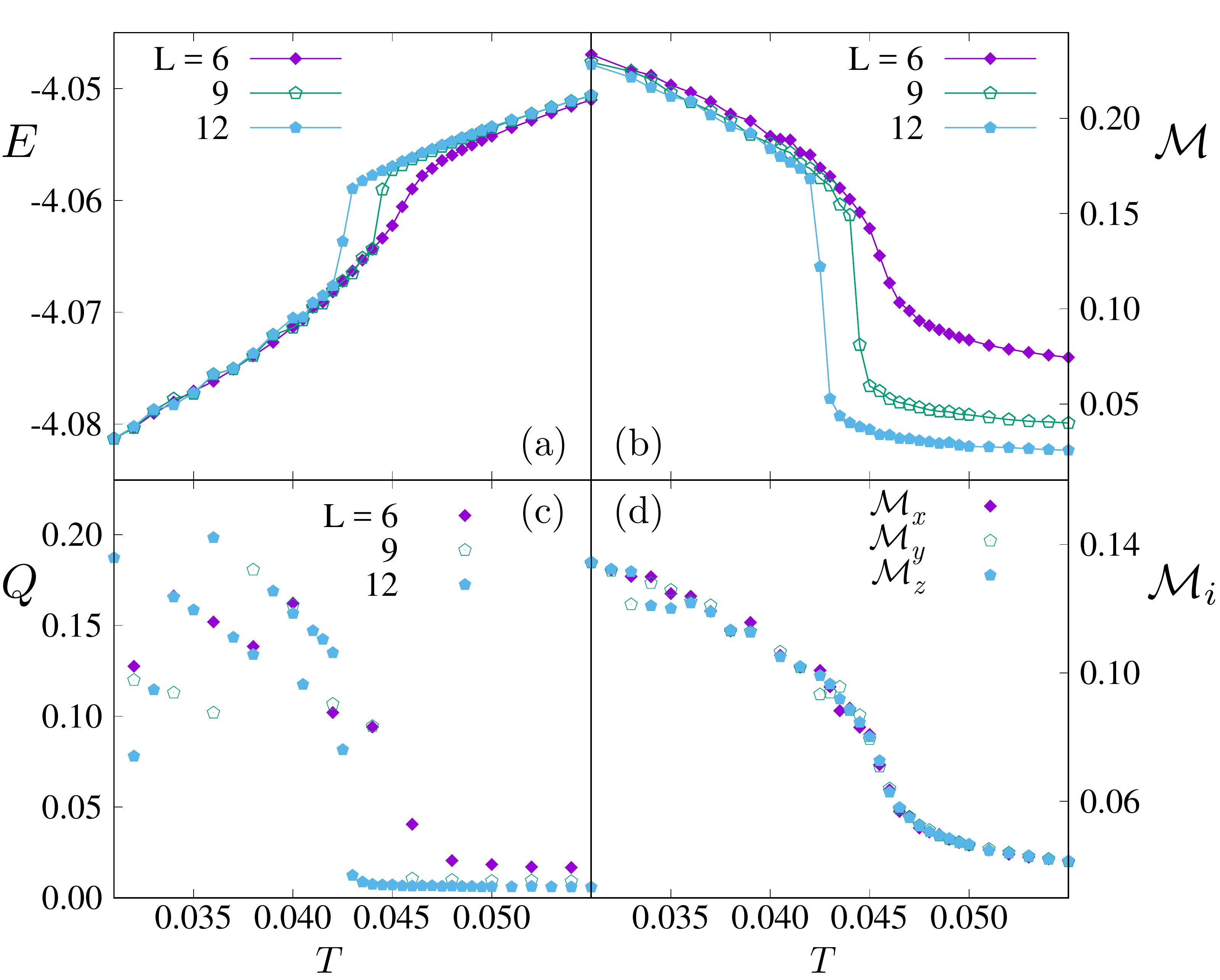}
\caption{
\label{fig:MC1}
(Color online) The temperature dependence of some quantities for the 2/3-filled coupled Kondo chains with $J = 1$. (a) the energy density and (b) the magnetic order parameter $\mathcal M$ shows the first order phase transition. (c) the nematic order parameter $Q$ bifurcates into multiple branches below the phase transition point. (d) the partial magnetic order parameter, $\mathcal M_x, \mathcal M_y, \mathcal M_z$, which are summation of $\bm\Phi_m$ at wavevectors $\mathbf q_m = \left(1, \pm\frac{1}{3}, \pm\frac{1}{3} \right)$, $ \left(\pm\frac{1}{3}, 1, \pm\frac{1}{3} \right) $, $\left(\pm\frac{1}{3},\pm\frac{1}{3},1 \right)$, respectively, with $L=6$ show no significant difference at low temperatures.
}
\end{center}
\end{figure}

To characterize the complex multiple-$\mathbf q$ magnetic order in the $\mathbf q =(\frac{1}{3},\frac{1}{3}, 1)$ phase, we introduce vector order parameters $\bm\Phi_{m} \equiv (1/N) \sum_j \mathbf S_j \,\exp(i \mathbf q_m\cdot \mathbf r_j)$, which are the Fourier modes of spins at the 12 symmetry-related wavevectors $\mathbf q_m = \left(\pm\frac{1}{3},\pm\frac{1}{3},1 \right) $, $ \left(\pm\frac{1}{3}, 1, \pm\frac{1}{3} \right) $, and $\left(1, \pm\frac{1}{3}, \pm\frac{1}{3} \right)$. Phenomenologically, the phase transition is described by a Landau free energy expansion~\cite{reimers91}
\begin{eqnarray}
	\mathcal{F} &=& \alpha (T - T_c) \sum_{m} |\bm\Phi_m|^2 + \beta \sum_{m} |\bm\Phi_{m}|^4  \\
	& &  + \sum_{m,n,k,l}\!\!\!\!' \, \,\, \lambda_{mnkl} (\bm\Phi_m\cdot\bm\Phi_n)(\bm\Phi_k \cdot \bm\Phi_l) + \cdots, \nonumber
\end{eqnarray}
where $\alpha, \beta > 0$, and the prime in the summation indicates the condition of momentum conservation, i.e. $\mathbf q_m + \mathbf q_n + \mathbf q_k + \mathbf q_l = 0$ module a reciprocal lattice vector. The overall $\mathbf q =(\frac{1}{3},\frac{1}{3}, 1)$ magnetic ordering is measured by the order parameter
\begin{eqnarray}
	\mathcal{M} = \left(\sum_{m=1}^{12} |\bm\Phi_m|^2 \right)^{1/2}.
\end{eqnarray} 
The temperature dependence of the $\mathcal{M}$, shown in Fig.~\ref{fig:MC1}(b), clearly indicates that these vector order parameters develop a nonzero expectation value at $T < T_c$, where $T_c$ is estimated to be $0.045t$ for $J=t$. Detailed structure of this $\mathbf q = (\frac{1}{3}, \frac{1}{3}, 1)$ magnetic order is determined by the interaction terms $\lambda_{mnkl}$, which are very difficult to compute analytically. Our extensive Monte Carlo simulations, on the other hand, seem to observe a multitude of different magnetic structures and a possible glassy regime below~$T_c$.

\begin{figure}[b]
\begin{center}
\includegraphics[width = 0.99\columnwidth]{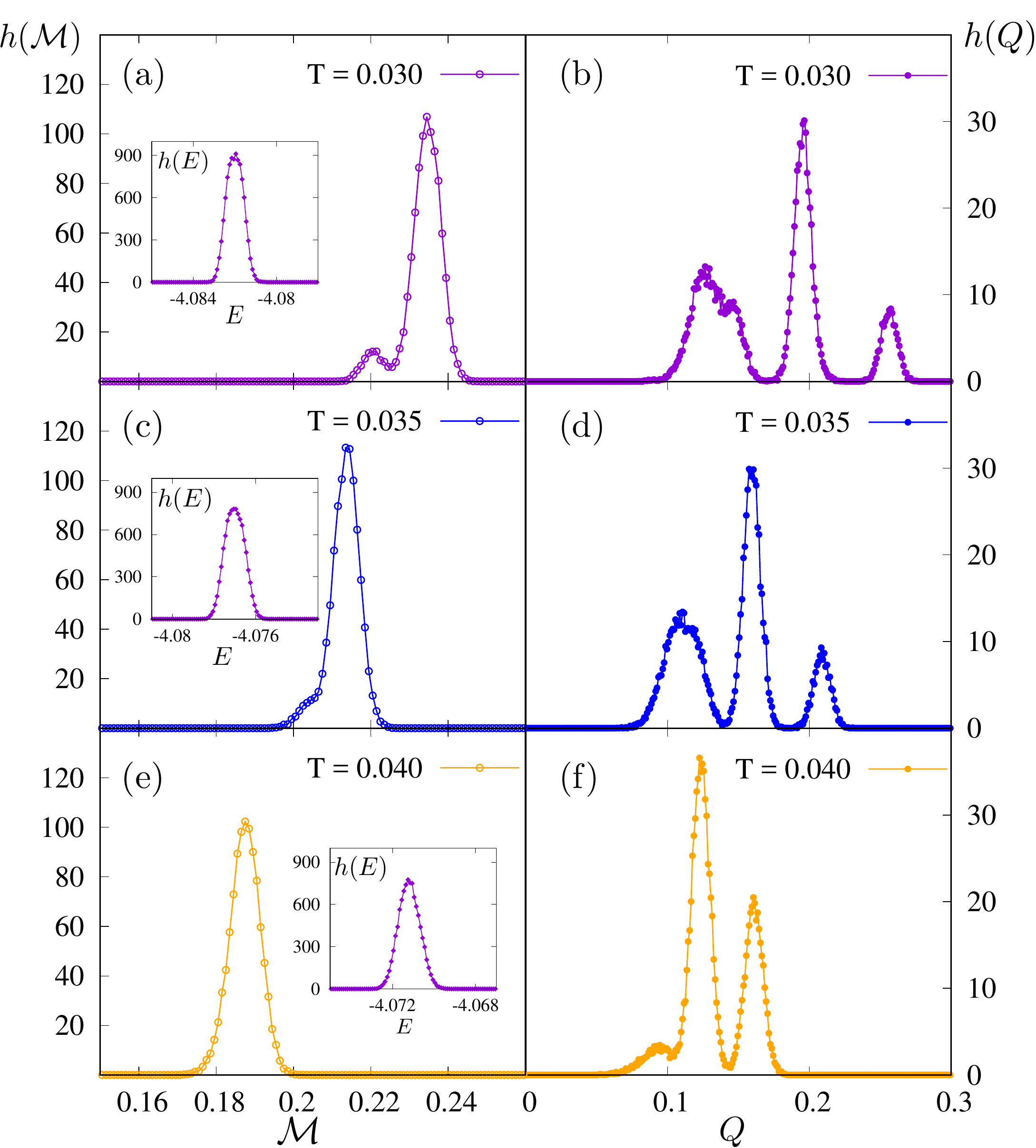}
\caption{
\label{fig:histogram}
(Color online) Probability distribution function for the magnetic order parameter $\mathcal{M}$, the nematic order parameter $Q$ and the energy density $E$ (insets) at three different temperatures below $T_c$. These curves are obtained from extensive Monte Carlo simulations on lattices of $J=1, L = 9$.}
\end{center}
\end{figure}

To explore this intriguing glassy phase, we compute the so-called nematic order-parameter $Q$ for spin structures obtained from our simulations. Essentially, this order parameter provides a measure of the collinearity of spins. It is given by the largest eigenvalue of the traceless matrix $\mathcal{Q}_{\mu\nu} \equiv \langle S_\mu \,S_\nu - \delta_{\mu\nu}/3 \rangle$ ($\mu, \nu = x, y, z$)~\cite{chaikin95}. Interestingly, the temperature dependence of the nematic order, shown in Fig.~\ref{fig:MC1}(c), exhibits three branches below the critical temperature $T_c$, implying distinct configurations of the $\mathbf q=(\frac{1}{3}, \frac{1}{3}, 1)$ magnetic order. To demonstrate this quasi-degeneracy directly, Fig.~\ref{fig:histogram}~shows the probability distribution of energy density $E$, magnetic order parameter $\mathcal{M}$, and spin nematic order parameter $Q$ at three different temperatures below $T_c$. Interestingly, while a single prominent peak is observed in the distribution of energy and magnetic order, the histogram of the nematic order parameter $Q$ exhibits several peaks, consistent with the multiple branches in Fig.~\ref{fig:MC1}(c). This finding clearly indicates a quasi degeneracy of the multiple-$\mathbf q$ magnetic orders, and the various quasi-degenerate $\mathbf q = (\frac{1}{3},\frac{1}{3},1)$ structures can be divided into three different groups according to their collinearity. We note that a systematic finite-size study is required in order to see whether this quasi-degeneracy structure persists in the thermodynamic limit. However, due to the limitation of our current Monte Carlo simulations that is based on the exact diagonalization method, it is already too costly to compute the histogram for $L=12$ lattices. Nonetheless, we have compared the histograms of $L = 6$ and $L = 9$ systems and found similar results. In fact, the multiple-peak feature is even more pronounced in the $L = 9$ histogram than the $L = 6$ one.

\begin{figure}[t]
\begin{center}
\includegraphics[width = 0.8\columnwidth]{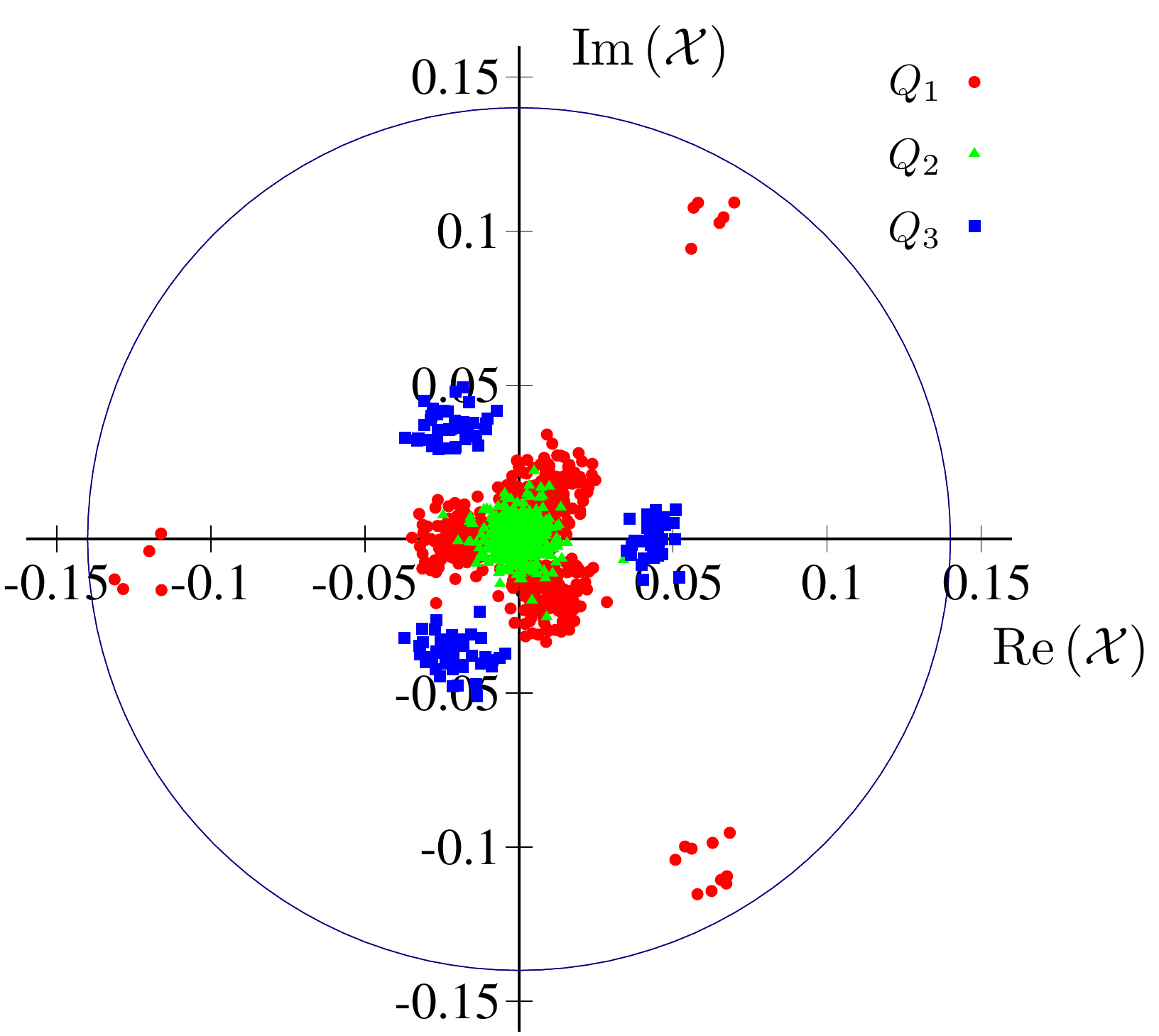}
\caption{
\label{fig:Rq}
(Color online) Distribution for $\mathcal{X}$ in the complex plane. Red, green, blue points represent independent samples whose nematic order parameter $Q$ is in the left, middle and right peaks, respectively, of the histogram $h(Q)$ in Fig.~\ref{fig:histogram}(b). Namely, $Q_1 \in \left[0,0.17\right)$, $Q_2 \in \left[0.17,0.23\right)$, $Q_3 \in \left[0.23,0.3\right]$. The figure is obtained with 1000 samples of the system at $T=0.03, J=1, L=6$.}
\end{center}
\end{figure}

Another important question is whether the cubic symmetry remains in the $\mathbf q = (\frac{1}{3},\frac{1}{3}, 1)$ magnetically ordered phase. To answer this question, we first define the partial magnetic order parameters $\mathcal M_x$, $\mathcal M_y$, and $\mathcal M_z$, which are sum of $|\bm\Phi_m|^2$ at wavevectors $\mathbf q_m = \left(1, \pm\frac{1}{3}, \pm\frac{1}{3} \right)$, $ \left(\pm\frac{1}{3}, 1, \pm\frac{1}{3} \right) $, $\left(\pm\frac{1}{3},\pm\frac{1}{3},1 \right)$, respectively. The dependence of these partial magnetic orders are plotted in Fig.~\ref{fig:MC1}(d) as functions of temperature. It is apparent that the cubic symmetry in the low-$T$ phase is conserved {\em in average}. However, the issue remains whether individual multi-$\mathbf q$ configuration preserves the cubic symmetry. To this end, we define a complex order parameter 
\begin{eqnarray}
	\mathcal{X} = \mathcal{M}_x + \omega\mathcal{M}_y + \omega^2 \mathcal{M}_z
\end{eqnarray}
which measures the disparity between the three partial magnetic orders; here $\omega=e^{i\frac{2\pi}{3}}$. A symmetric phase with $\mathcal{M}_x \approx \mathcal{M}_y \approx \mathcal{M}_z$, thus gives rise to a vanishing complex order parameter $\mathcal{X} \approx 0$. Fig.~\ref{fig:Rq} shows the distribution of $\mathcal{X}$ obtained from 1000 independent Monte Carlo runs. Interestingly, we find strong correlation between the nematic order $Q$ and the cubic-symmetry parameter $\mathcal{X}$. Here magnetic orders belonging to distinct groups in the histogram (Fig.~\ref{fig:histogram}) are labeled by three different colors. For example, the $\mathcal{X}$ parameters corresponding to the middle peak of $h(Q)$ in Fig.~\ref{fig:histogram}(b) cluster around the origin, indicating that these $\mathbf q = (\frac{1}{3}, \frac{1}{3}, 1)$ magnetic orders approximately preserves the cubic symmetry. The two distinct parts with smaller $Q$ illustrate the possible existence of two phases corresponding to this peak. On the other hand, magnetic orders with large $Q$ tends to break the cubic symmetry. However, it is worth noting that the cubic symmetry is recovered when averaging over multiple domains each characterized by a different $\mathcal{X}$ in the system. This picture of quasi-degenerate multi-$\mathbf q$ manifold is thus consistent with the experimental observation that GeFe$_2$O$_4$ retains cubic symmetry in the low-$T$ magnetic glassy phase. 

\begin{figure}[t]
\begin{center}
\includegraphics[width = 0.8\columnwidth]{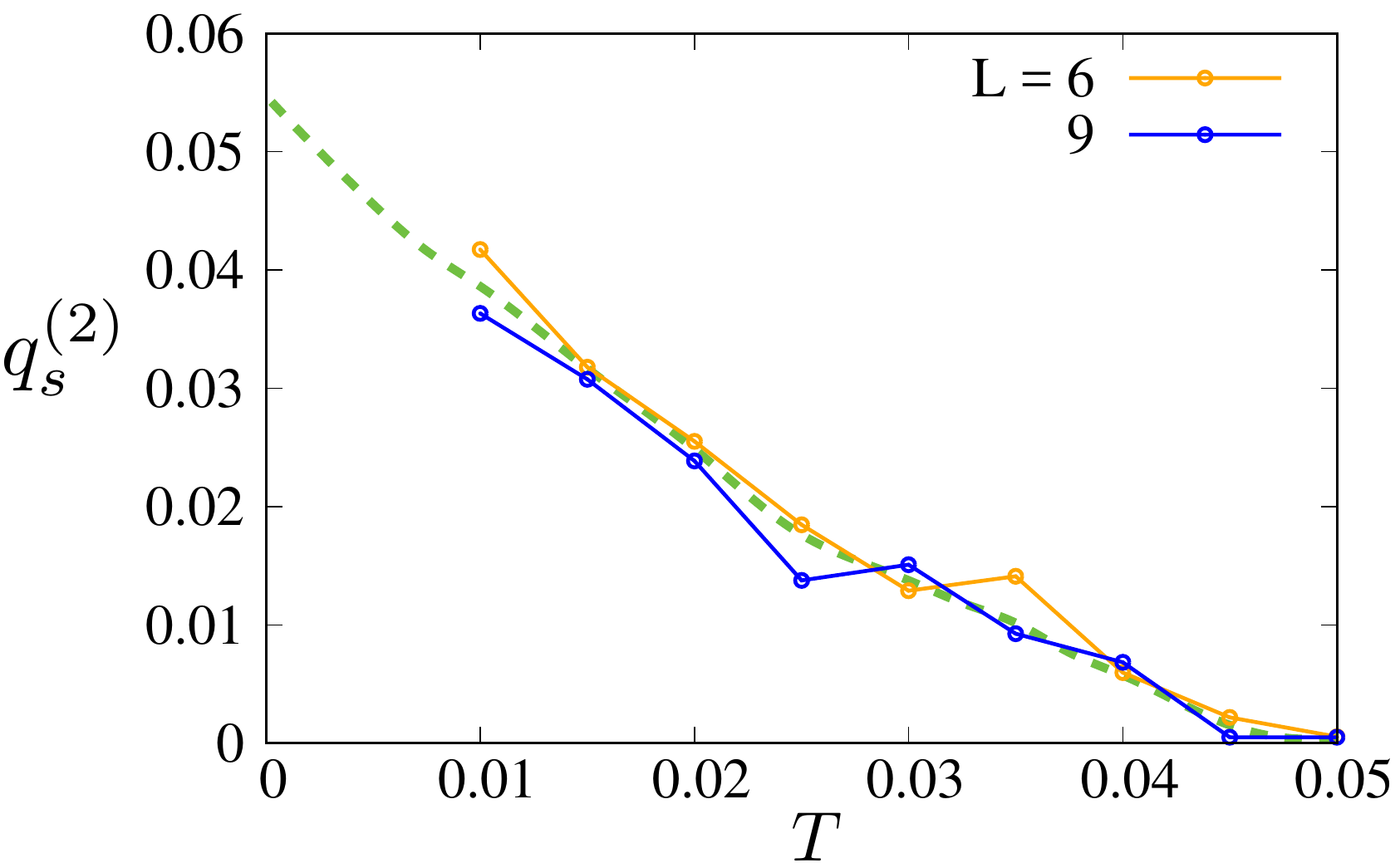}
\caption{
\label{fig:qs2}
(Color online) The spin freezing parameter $q_s^{(2)}$ as a function of temperature with the electron-spin coupling $J = 1$. The green dashed line is a guide to the eye. The distribution of $q_s^{(2)}$ is rather asymmetric and non-Gaussian.}
\end{center}
\end{figure}

We also compute the spin freezing parameter defined as $q^{(2)}_{\rm SG} = \sum_{\mu\nu} \langle q_{\mu\nu}^2 \rangle$~\cite{viet09}, where $q_{\mu\nu}  = (1/N) \sum_i S^{(a)}_{i,\mu} S^{(b)}_{i, \nu}$ denotes the overlap of spins obtained from two replicas $a$ and $b$. This parameter is nonzero when spins are frozen either in an ordered or a random configuration. Fig.~\ref{fig:qs2} shows the temperature dependence of the $q^{(2)}_{\rm SG}$ parameter computed from our Monte Carlo simulations for $J = 1$. The freezing parameter starts to grow at the magnetic transition point. Moreover, the curves for different lattice sizes show rather weak finite size dependence, consistent with a first-order phase transition scenario. Extrapolating to zero temperature, we obtain a nonzero, yet rather small $q^{(2)}_{\rm SG} \approx 0.05$. This near vanishing of the freezing parameter can be attributed to the quasi-degeneracy of the multiple-$\mathbf q$ manifold of the $(\frac{1}{3},\frac{1}{3}, 1)$ phase. Similar multiple-$\mathbf q$ glassy states have also been observed in $J_1$-$J_2$ Heisenberg pyrochlore antiferromagnets~\cite{chern08,okubo11}.

\section{Conclusion and outlook}
 
To summarize, we have presented a thorough numerical study of a new type of itinerant frustrated magnetism on the pyrochlore lattice. In this model, the pyrochlore magnet can be viewed as a cross-linking network of Kondo chains. We have obtained the phase diagrams at two representative filling fractions $n =1/2$ and $2/3$. This model provides a natural explanation to complex spin and orbital structures observed in several spinels compound, which are very difficult to understand within localized spin models. 
Importantly, this magnetic phase provides a rather consistent explanation for the recently observed magnetic order in spinel GeFe$_2$O$_4$~\cite{zhu16}. In this compound, two of the 6 $d$-electrons of the magnetic Fe$^{2+}$ ion occupy the $e_g$ level, forming the local spins $\{\mathbf S_i\}$ with length $S = 1$. The other 4 $d$-electrons partially fill the $t_{2g}$ orbitals, forming quasi-1D tight-binding chains with a filling fraction $n = 2/3$. Instead of a sharp Bragg peak, neutron scattering experiments show diffusive peaks centered at $\mathbf q = (\frac{1}{3}, \frac{1}{3}, 1)$ wavevectors, indicating a short-range spin ordering in this material. This observation is also consistent with the glassy $(\frac{1}{3},\frac{1}{3},1)$ phase of our model. Since the magnetic transition is first order, the correlation length remains finite throughout the phase transition. The large quasi-degeneracy of spin orders in this phase also means that most likely the low temperature phase of GeFe$_2$O$_4$ consists of finite domains of different magnetic structures. 

The $\mathbf q = (\frac{1}{3},\frac{1}{3},1)$ glassy phase is reminiscent to other magnetic glassy states reported in strongly correlated systems, including frustrated magnets~\cite{gingras97,samarakoon16}, high-$T_c$ superconducting materials~\cite{matsuda00,katayama10}, and spin-orbital Mott insulator~\cite{luo13}. All these states are characterized by diffuse scattering at well defined wavevectors, indicating the short-range nature of magnetic orders. A plausible picture for these glassy magnets is the coexistence of domains with different spin structures separated by domain-walls. Moreover, they also exhibit dynamical behaviors~\cite{halperin77,mross15,samarakoon17} that are different from conventional spin glass. Our work along with previous studies~\cite{chern08,okubo11} suggest that multiple-$\mathbf q$ magnetic ordering in frustrated magnets provides a new route to realize such unconventional glassy magnets and GeFe$_2$O$_4$ is a potential candidate.

\bigskip

{\em Acknowledgement}. We thank Xianglin Ke for sharing us the unpublished experimental data and several insightful discussions.

\appendix

\section{1D Kondo chains}
\label{sec:1D}

In this section, we consider the ground state of 1D Kondo chains, which are the backbone of the itinerant frustrated model on the pyrochlore lattice discussed in the main text. The Hamilton of a Kondo chain is 
\begin{eqnarray}
	\label{eq:H_DE2}
	\mathcal{H} = -t \sum_{i}\sum_{\sigma = \uparrow, \downarrow} \left( c^{\dagger}_{i, \sigma} \,c^{\;}_{i+1, \sigma} + {\rm h.c.}\right)
	- J \sum_i \mathbf S_i \cdot \mathbf s_{i}, \qquad
\end{eqnarray}
where $c^\dagger_{i, \sigma}$ is the creation operator of electrons at site-$i$ with spin $\sigma$, $t$ is the nearest-neighbor hopping constant, $J$ is the Hund's coupling strength, $\mathbf S_i$ is local magnetic moment, and $\mathbf s_{i} = \sum_{\alpha,\beta} c^{\dagger}_{i\alpha}  {\bm \sigma_{\alpha\beta}} c_{i\beta}$ is the spin of the conduction electron. Since we are interested in magnetically ordered or glass states with frozen nonzero moments, we further assume $\mathbf S_i$ are classical spins with magnitude $|\mathbf S_i| = 1$.
The zero-temperature phase diagram of the classical 1D Kondo chain in the $\mu$-$J$ plane, where $\mu$ is the chemical potential of the electrons, has been mapped out in Ref.~\cite{SI_Kawamura15}. Here, instead, we focus on the Kondo chain with a fixed filling fraction $n = 1/2$ and $n = 1/3$, and obtain the ground states as a function of $J$. Due to particle-hole symmetry, the $\frac{2}{3}$ filling case studied in the main text for the 3D pyrochlore model, is equivalent to the $\frac{1}{3}$ filling case.  


We perform extensive Monte Carlo simulations with Metropolis algorithm to obtain the ground states of the 1D Knodo chain. While most of the results discussed below were obtained from the chain with $N = 72$ spins, we have also conducted simulations with different chain lengths and boundary conditions (periodic vs open boundary conditions) in order to eliminate the finite size effects. To avoid freezing problems, we started our simulations at a relatively high temperature and perform annealing simulations by slowly reducing the temperature. The final spin configuration is determined at $T\approx 10^{-8}$. The structure factor $S(q)$ and correlation function are evaluated and averaged at the final temperature.


We first discuss the half-filling case. Our simulations find a ground state with N\'eel order, i.e. $\uparrow\downarrow\uparrow\downarrow \cdots$, for all values of electron-spin coupling $J$, consistent with the results obtained in Ref.~\cite{SI_Kawamura15}. One can understand the stabilization of the N\'eel order from the weak as well as the strong coupling limits. In the small $J$ limit, the nesting of the Fermi points of a half-filled chain leads to a weak-coupling instability with respect to perturbation of N\'eel wavevector $q = 2 k_F = \pi$. The energy of the N\'eel ordered state is lowered by opening a spectral gap at the Fermi points. Furthermore, our simulation finds that the energy-gain is maximized by collinear N\'eel order.  In the opposite large $J$ regime, the half-filled chain is a special case in the sense that there exists a macroscopic degeneracy in the $J \to \infty$ limit. In this strong coupling limit, each site binds an electron whose spin is aligned with the local moment $\mathbf S_i$, whose direction can point in an arbitrary direction. As discussed in the main text, this huge degeneracy is lifted by the nearest-neighbor hopping, giving rise to an effective {\em antiferromagnetic} spin-spin interaction $J_{\rm AF} \mathbf S_i \cdot \mathbf S_j$, where $J_{\rm AF} \sim t^2/J > 0$. Consequently, the N\'eel order is also stabilized in this large $J$ limit. The cross-linking geometry in the pyorhclore lattice leads to geometrical frustration of N\'eel ordered chains. The system ends up in an all-in-all-out long-range order in which the N\'eel order coexists with a ferromagnetic component in each chain.


\begin{figure*}[htbp]
\begin{center}
\includegraphics[width = 1.5\columnwidth]{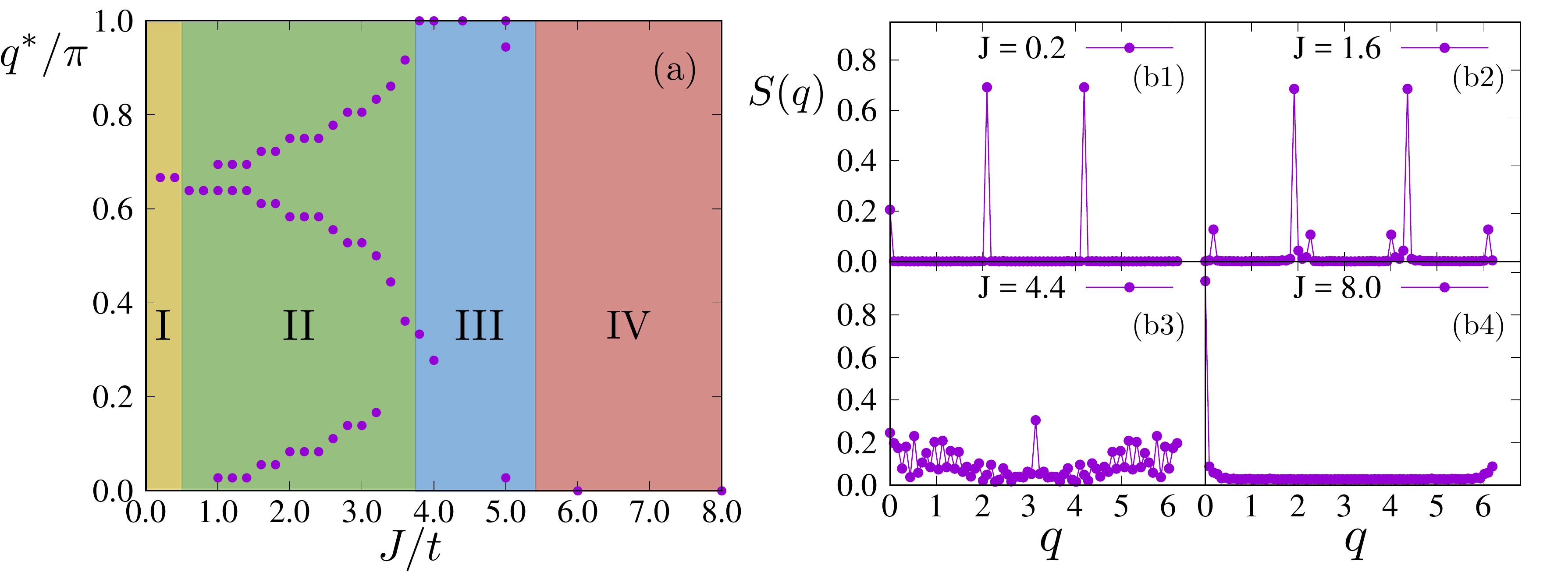}
\caption{\label{fig:phase} (Color online) (a): Wave vectors $q^{*}$ where the module of $\mathbf S(q)$ reaches local maximum for $\frac{1}{3}$ filling. (b1) $\sim$ (b4): $S(q)$ as a function of $q$ for characteristic $J=0.2,1.6,4.4,8.0$ in region I, II, III, IV.}
\end{center}
\end{figure*}

The $1/3$-filled Kondo chain displays a richer phase diagram as shown in Fig.~\ref{fig:phase}(a). Here we plot the wavevector $q^*$, which corresponds to the maxima of $S(q)$, as a function of $J$. At $J \lesssim 0.5$, the ground state shows a spin configuration with a period of 3, represented by a wavevector at $q^{*} = \frac{2}{3}\pi$, as shown in Fig. \ref{fig:phase} (b1). Again, this magnetic order arises from the weak-coupling instability due to the Fermi point nesting $q^* = 2 k_F$ for a 1/3-filled chain. The wavevector $q^* = 2\pi/3$ bifurcates at $J\approx 0.5$ with one branch gradually going down and the other one rising up to $\frac{1}{2}\pi$ (see Fig. \ref{fig:phase} (b2)). The small plateaus for $q^*$ may result from the finite size effect. In region III, the spin structure tends to be non-coplanar and rather complicated, represented by a less pronounced peak at $q^*=\pi$ (see Fig. \ref{fig:phase} (b3)). Starting from $J\approx 5.5$, the ground state is ferromagnetic (Fig. \ref{fig:phase} (b4)). The 3-period phase at small $J$ agrees with that of 3D pyrochlore lattice, while at intermediate $J$, the gradual change of $q^*$ is broken by the 3D structure and replaced by a $(1/2,1/2,1/2)$ order. In both 1D and 3D models, a large $J$ gives rise to the ferromagnetic phase. The evolution of the most pronounced wave vector, which is the line in the middle in region II of Fig. \ref{fig:phase}(a), shows the same trend as that of the quantum Kondo chain~\cite{SI_Garcia}.

\begin{figure*}[]
\begin{center}
\includegraphics[scale=0.35]{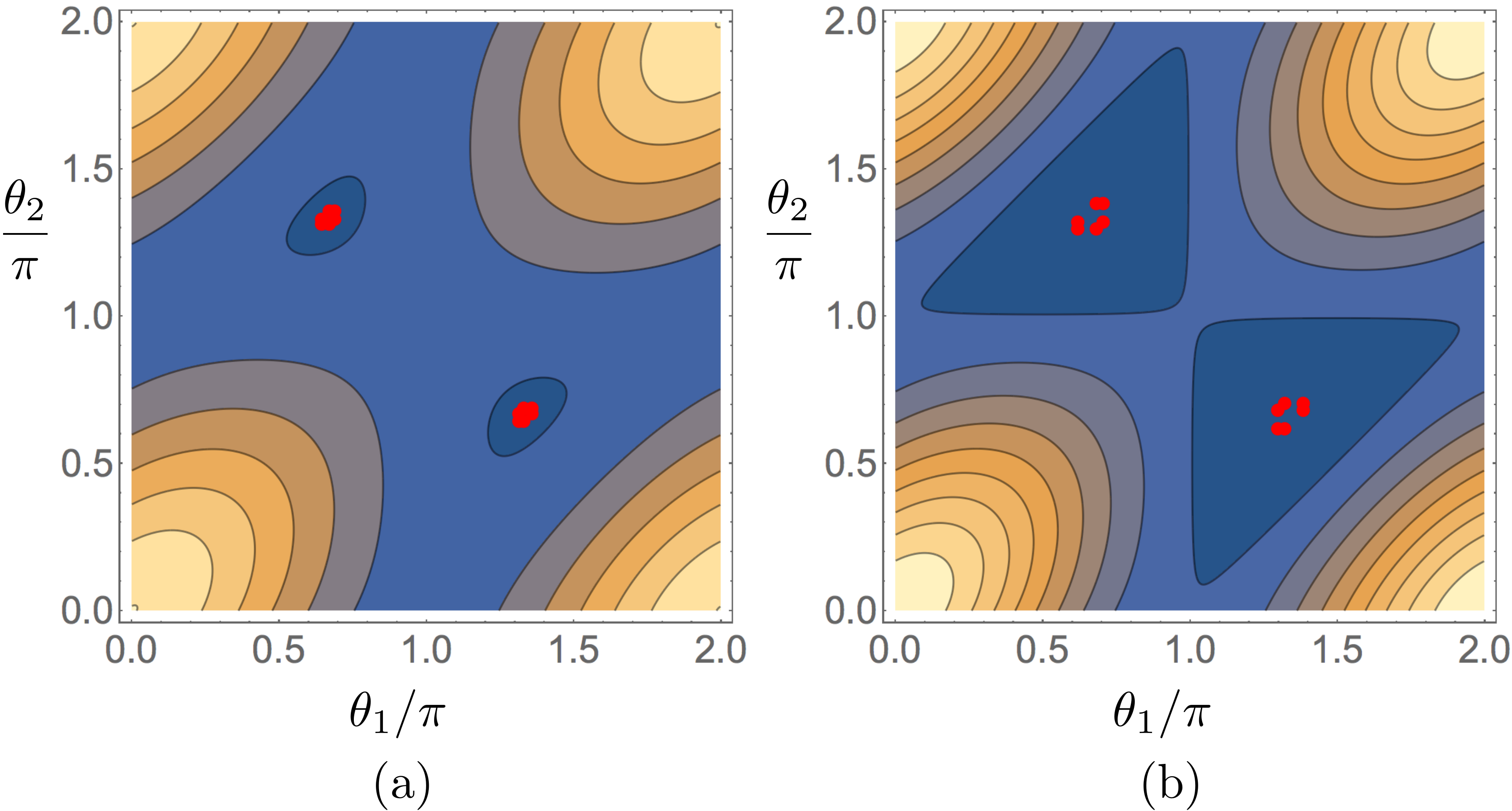}
\caption{(Color online) Energy contour plot with respect to $\theta_1$ and $\theta_2$ for $\left(a\right)J=0.2, \left(b\right) J=0.4$. The red dots indicate the optimal configuration. The range for $\theta_1$ and $\theta_2$ is $\left[0,2\pi\right]$. The 12 dots represent the same or symmetry related configuration. For $J=0.2$, the ground state is the state such that the angles between each of the 3 pairs of  $\mathbf S_0, \mathbf S_1, \mathbf S_2$ are $115.8^{\circ}, 120.6^{\circ}, 123.6^{\circ}$. For $J=0.4$, they are $111^{\circ}, 112.4^{\circ}, 126.6^{\circ}$.}
\end{center}
\end{figure*}

Here we identify the period-3 ground state for small $J=0.2, 0.4$ at $1/3$ filling. Since we expect the ground state with a 3-period structure due to the Fermi point nesting mechanism as indicated by the Monte Carlo simulations, we can then Fourier transform the real space Hamiltonian to a k-space Hamiltonian $ \mathcal {H} = \sum_k \sum_{i,j=0}^{2} c_{i\alpha}^{\dagger } (k) H_{i\alpha,j\beta} (k) c_{j\beta} (k) $ in which
\begin{eqnarray}
  & H(k) ={\left( \begin{matrix}
   -\frac{1}{2}J{{{\bm{\sigma }}}}\cdot {{{\mathbf{S}}}_{0}} & -t{{e}^{ik}}\sigma_0 & -t{{e}^{-ik}} \sigma_0  \\
   -t{{e}^{-ik}}\sigma_0 & -\frac{1}{2}J{{{\bm{\sigma }}}}\cdot {{{\mathbf{S}}}_{1}} & -t{{e}^{ik}}\sigma_0  \\
   -t{{e}^{ik}}\sigma_0 & -t{{e}^{-ik}}\sigma_0 & -\frac{1}{2}J{{{\bm{\sigma }}}}\cdot {{{\mathbf{S}}}_{2}}  \\
\end{matrix} \right)} 
\end{eqnarray}
where $\sigma_0$ is the $2 \times 2$ identity matrix. Working on the k-space, we try to identify the lowest energy local spin configuration. The whole spin chain is composed of multiple periodic duplicates of the first three spins $\mathbf S_0, \mathbf S_1, \mathbf S_2$. It is convenient to set $\mathbf S_0 = (0,0,1), \mathbf S_1=(\sin{\theta_1},0,\cos{\theta_1}), \mathbf S_2= (\sin{\theta_2}\cos{\phi_2},\sin{\theta_2}\sin{\phi_2},\cos{\theta_2})$. Scanning over $\theta_1,\theta_2,\phi_2$ shows the minimum energy is obtained when $\mathbf S_0, \mathbf S_1$ and $\mathbf S_2$ are coplanar, which allows us to  set $\phi_2=0$. We can then scan over $\theta_1,\theta_2$ only. The final optimal configurations with $J=0.2,0.4$ are presented in the figure below. Although it is tempted to consider that the structure with the angle between any pair of spins being $120^{\circ}$ is the best configuration, our results show that the optimal state is close to but not exactly the $120^{\circ}$ structure and varies with $J$.

\end{document}